\documentclass[12pt]{iopart}
\usepackage{amssymb}
\usepackage{epsfig}
\usepackage{subfigure}
\usepackage{graphicx}
\usepackage{textcomp}
\usepackage{cite}
\usepackage{float}
\usepackage{subfigure}

\expandafter\let\csname equation*\endcsname\relax
\expandafter\let\csname endequation*\endcsname\relax
\usepackage{amsmath}
\newcommand{\agt}{\mathrel{\raise.3ex\hbox{$>$\kern-.75em\lower1ex\hbox{$\sim$}}}}

\begin{document}
\title{Large fluctuations in diffusion-controlled absorption}
\author{Baruch Meerson}
\address{Racah Institute of Physics, Hebrew University of
Jerusalem, Jerusalem 91904, Israel}
\author{S. Redner} \address{Department of Physics, Boston University, Boston,
  MA 02215, USA}
  \address{Santa Fe Institute, 1399 Hyde Park Road, Santa Fe, New
  Mexico 87501, USA}

\begin{abstract}

  Suppose that $N_0$ independently diffusing particles, each with diffusivity
  $D$, are initially released at $x=\ell>0$ on the semi-infinite interval
  $0\leq x<\infty$ with an absorber at $x=0$.  We determine the probability
  ${\cal P}(N)$ that $N$ particles survive until time $t=T$.  We also employ
  macroscopic fluctuation theory to find the most likely history of the
  system, conditional on there being exactly $N$ survivors at time $t=T$.
  Depending on the basic parameter $\ell/\sqrt{4DT}$, very different histories
  can contribute to the extreme cases of $N=N_0$ (all
  particles survive) and $N=0$ (no survivors).  For large values of $\ell/\sqrt{4DT}$,
  the leading contribution to ${\cal P}(N=0)$
  comes from an effective point-like quasiparticle that contains all the
  $N_0$ particles and moves ballistically toward the absorber until
  absorption occurs.

\end{abstract}
\maketitle

\section{Problem Statement}

When $N_0$ independent diffusing particles are released at $x=\ell>0$ on the
infinite half-line $x>0$ with the origin being absorbing, a basic
characteristic of the dynamics is the number of surviving particles.  In this
work, we focus on the \emph{distribution} of the number of survivors at a
specified time.  As we shall discuss, disparate histories, that are very
different from typical histories of diffusion, can contribute to the extreme cases where
(a) all particles survive or (b) none survive.

The probability that a single particle survives up to time $t$ is
\begin{equation}\label{expectedS}
  \overline{\Theta}(t)=\int_0^{\infty} c(x,t) \,dx\,,
\end{equation}
where $c(x,t)$ is the probability density for the particle to be at position
$x$ at time $t$~\cite{W,Redner}.  This density obeys the diffusion equation
\begin{equation}
\partial_t c =D \,\partial_{x}^2 c\,,
\label{difeq}
\end{equation}
whose solution, subject to the initial condition $c(x,0)=\delta(x\!-\!\ell)$ and
the absorbing boundary condition $c(0,t)=0$, is
\begin{equation}\label{mf1}
c(x,t)=\frac{1}{\sqrt{4\pi D t}}\left[e^{-\frac{(x-\ell)^2}{4Dt}}-e^{-\frac{(x+\ell)^2}{4Dt}}\right].
\end{equation}
Then Eq.~(\ref{expectedS}) gives
\begin{equation}\label{expectedSresult}
    \overline{\Theta}(T)=\text{erf}\left(\frac{\ell}{\sqrt{4 D T}}\right),
\end{equation}
where $\text{erf}(z)=(2/\sqrt{\pi}) \int_0^z e^{-u^2}\,du$ is the error
function.  Since the particles are non-interacting, the average number
$\overline{N}(T)$ of survivors at $t=T$, when there are $N_0$ particles
initially at $x=\ell$, is
\begin{equation}\label{barN}
\overline{N}(T)= N_0 \,\overline{\Theta}(T)=N_0 \;\text{erf}\left(\frac{\ell}{\sqrt{4 D T}}\right).
\end{equation}
Our goal is to determine the probability ${\cal P}(N)$ to observe \emph{any}
number of surviving particles $0\leq N\leq N_0$ at a specified observation
time $t=T$. We shall also seek the most likely density history of the system,
conditional on the survival of exactly $N$ particles at $t=T$.  The number of
survivors can be equal to $N_0$ (all particles survive until $t=T$) or to $0$
(no survivors). As we shall see, the relevant histories can be quite unusual
in these cases and depend on the basic parameter $\ell/\sqrt{4DT}$.

Since the particles are independent, the probability ${\cal P}(N)$ can be
found exactly from a microscopic theory, as given in the next section.  The
most likely history of the system is harder to find from microscopic
arguments, even for independent particles.  This history can be readily
found, however, within the framework of the approximate \emph{macroscopic
  fluctuation theory} of Bertini et al.~\cite{Bertini,Jona}, which identifies
the typical number of particles in the relevant region of space as the large
parameter.  As will be presented in Sec.~\ref{MFT}, this most likely density
history provides fascinating insights into the nature of large fluctuations
in diffusion-controlled absorption.

\section{${\cal P}(N)$ and its limiting behaviors}
\label{P(N)}

\begin{figure}[h]
\subfigure[]{\includegraphics[width=0.33\textwidth,clip=]{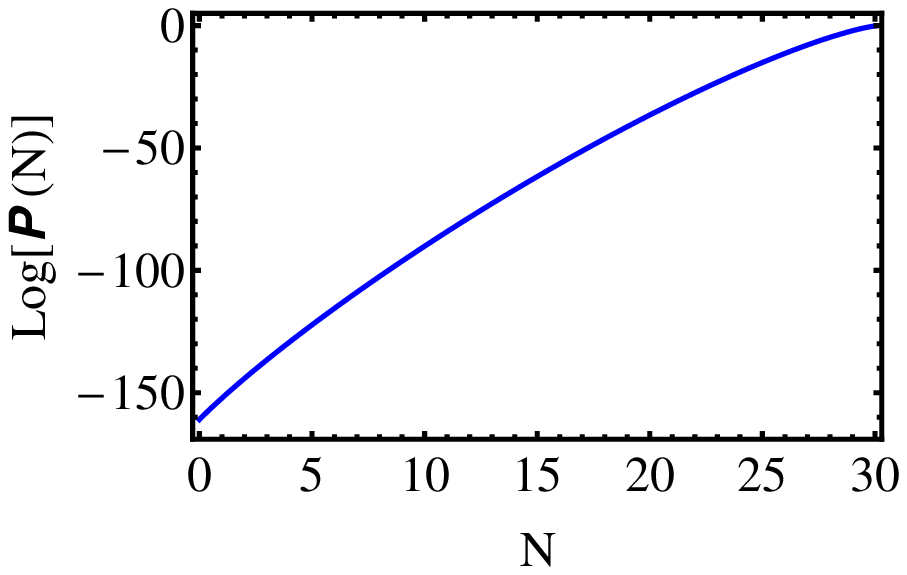}}
\subfigure[]{\includegraphics[width=0.33\textwidth,clip=]{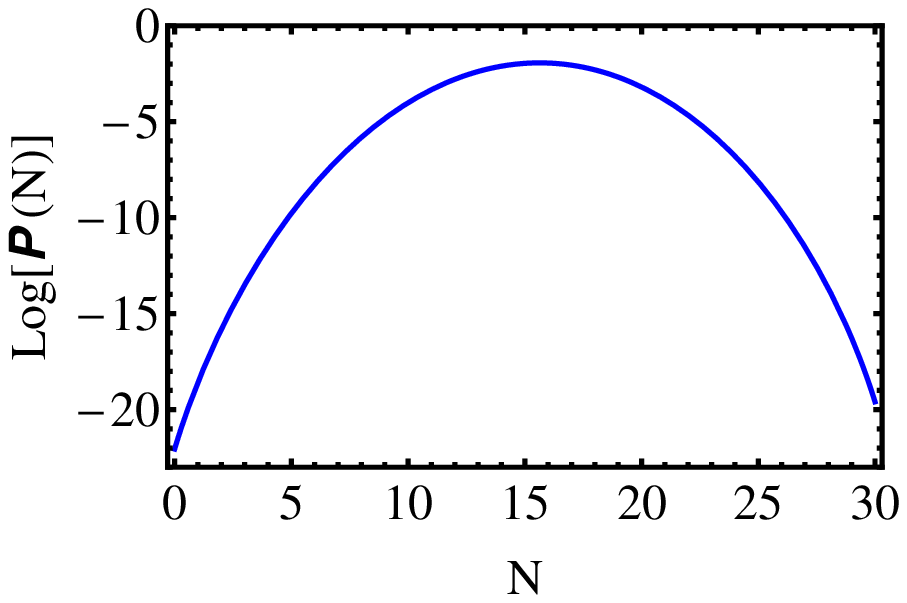}}
\subfigure[]{\includegraphics[width=0.33\textwidth,clip=]{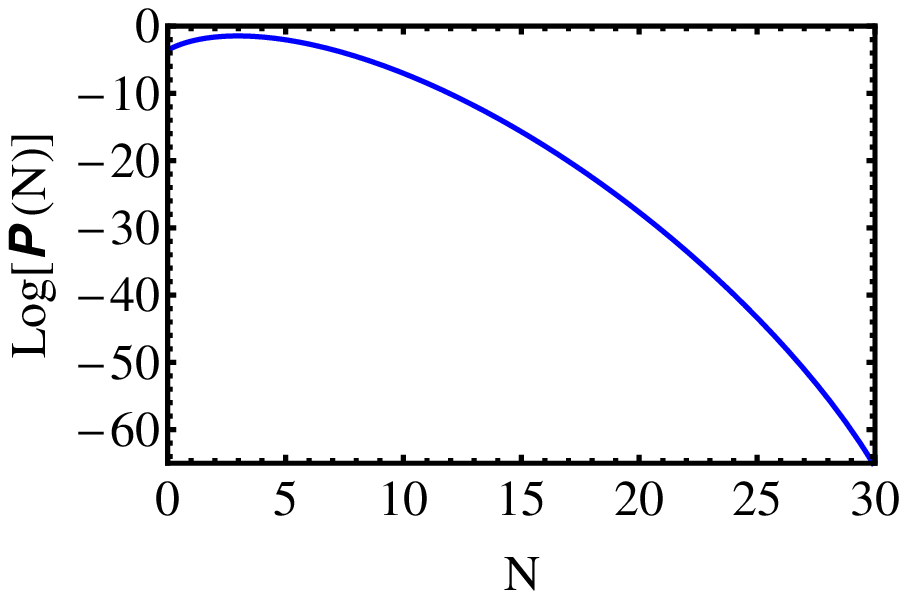}}
\caption{Logarithm of the survival probability ${\cal P}$ versus $N$ for
  $N_0=30$ and three values of $\ell/\sqrt{4 D T}$: (a) 2, (b) $1/2$ and (c)
  $1/10$.}
\label{P}
\end{figure}

The probability $\overline{\Theta}(T)$ that a single particle does not hit the
absorber by time $T$ (the survival probability) is given by
Eq.~\eqref{expectedSresult}.  The complementary probability that a single
particle hits the absorber by time $T$ is $1-\overline{\Theta}(T)$.  Since the
particles are independent, the probability that exactly $N$ out of $N_0$
particles survive up to time $T$ is given by the binomial distribution:
\begin{equation}
{\cal P}(N) = \binom{N_0}{N} \big[\overline{\Theta}(T)\big]^N \big[1-\overline{\Theta}(T)\big]^{N_0-N}.
\label{binom}
\end{equation}
In particular,
\begin{equation}\label{special}
{\cal P}(N\!=\!N_0)=\left[\overline{\Theta}(T)\right]^{N_0}\qquad\text{and}\qquad
{\cal P}(N\!=\!0)=\left[1-\overline{\Theta}(T)\right]^{N_0}.
\end{equation}
Figure \ref{P} shows $\ln {\cal P}$ versus $N$ for $N_0=30$ and three values
of the parameter $\ell/\sqrt{4 D T}$. For $\ell/\sqrt{4 D T}=2$,
corresponding to the situation where a particle typically has not yet
diffused to the origin, the survival probability is peaked at the initial
value $N=N_0=30$ and rapidly decreases as $N$ gets smaller.  In the
intermediate case of $\ell/\sqrt{4 D T}=1/2$, where a particle has
typically just diffused to the origin, the survival probability is peaked at
$N=16$ and nearly symmetric apart from the tails.  As one might expect, roughly one-half of the
particles have been absorbed at this point.  Finally, for
$\ell/\sqrt{4 D T}=1/10$, most of the particles have been absorbed.
Here, the survival probability is peaked at $N=3$ and rapidly decreases as
$N$ increases.

To elucidate the role of the parameter $\ell/\sqrt{4DT}$, we focus on two
extreme cases: $N=N_0$ (all particles survive until $t=T$) and $N=0$ (no
survivors at $t=T$), as well as the intermediate case with roughly similar
numbers of survivors and non-survivors, $N_0\gg 1$, $N\gg 1$ and $N_0-N\gg
1$.

\subsection{All particles survive until time $T$ ($N=N_0$)}

For $\ell/\sqrt{4DT} \gg 1$, we use the asymptotic
\begin{equation}\label{largez}
  \text{erf}\,z \simeq 1-\frac{e^{-z^2}}{\sqrt{\pi} z},\qquad z\gg 1\,,
\end{equation}
to give
\begin{equation}\label{1large}
\ln {\cal P}(N\!=\!N_0)\simeq - \frac{N_0 \sqrt{4DT}}{\sqrt{\pi} \ell}\,\,e^{-\frac{\ell^2}{4DT}}~.
\end{equation}
The right-hand side is extremely small so that ${\cal P}(N\!=\!N_0)$ is
extremely close to 1. This is expected as, for $\ell/\sqrt{4DT} \gg 1$, a particle typically travels a distance much smaller than $\ell$ during time $T$.     Conversely, when $\ell/\sqrt{4DT} \ll 1$,
we use the asymptotic $\text{erf}(z) \simeq (2/\sqrt{\pi})\,z$ to obtain
\begin{equation}\label{1small}
{\cal P}(N\!=\!N_0) \simeq \left(\frac{\ell}{\sqrt{\pi D T}}\right)^{N_0},
\end{equation}
which is very small, again as expected.

\subsection{No survivors at time $T$ ($N=0$)}

In the limit of $\ell/\sqrt{4DT} \gg 1$, we obtain
\begin{equation}\label{0large}
  \ln {\cal P}(N\!=\!0)\simeq - N_0 \left(\frac{\ell^2}{4DT} +\ln \frac{\sqrt{\pi} \ell}{\sqrt{4DT}}+ \dots \right),
\end{equation}
so the probability is very small as expected.  In the opposite limit of $\ell/\sqrt{4DT}
\ll 1$, we obtain
\begin{equation}\label{0small}
\ln {\cal P}(N\!=\!0)\simeq - N_0 \ell/\sqrt{\pi D T}.
\end{equation}
The resulting probability can be close to or much less than one, depending on
whether $N_0$ is small or large.

\subsection{Intermediate case}
\label{Stirling}
For the situation where $N_0\gg 1$, $N\gg 1$ and $N_0-N\gg 1$, we use
Stirling's approximation in Eq.~\eqref{binom} to obtain, after some algebra,
\begin{equation}
  -\ln {\cal P}(N) \simeq N_0\left[\Theta \ln \frac{\Theta}{\overline{\Theta}}+(1-\Theta)
    \ln \frac{1-\Theta}{1-\overline{\Theta}}\right]
  +\ln \left[\sqrt{2\pi N_0 \,\Theta (1-\Theta)}\right],
\label{Pappr}
\end{equation}
where $\Theta=N/N_0$ is the fraction of surviving particles. Note that the
survival probability $\overline{\Theta}=\overline{\Theta}(T)$, see
Eq.~(\ref{expectedSresult}), coincides with the expected (average) fraction
of surviving particles. For $N$ close to its expected value, that is, for $\Theta$ close to $\overline{\Theta}$,
the distribution ${\cal P}(N)$ is Gaussian:
\begin{equation}\label{gauss}
{\cal P}(N) \simeq \frac{1}{\sqrt{2\pi}\, \sigma}\,e^{-\frac{(N-\bar{N})^2}{2\sigma^2}}.
\end{equation}
The variance is $\sigma^2= N_0\, \overline{\Theta} (1-\overline{\Theta})$.
Interestingly, the fluctuations are maximal at an intermediate value of the parameter $\ell/\sqrt{4DT}$.
Indeed, the variance $\sigma^2$ vanishes at $\ell/\sqrt{4DT}\to 0$ and $\ell/\sqrt{4DT}\to \infty$ [see
Eq.~\eqref{expectedSresult}], and reaches a maximum value equal to  $\sigma^2=N_0/4$ when
$\overline{\Theta}=1/2$; that is, when $\ell/\sqrt{4DT}=0.4769\ldots$.

Notice that the first term on the right hand side of Eq.~(\ref{Pappr}), which
dominates $\ln {\cal P}(N)$, can be written in the scaling form
\begin{equation}\label{scaling}
-\ln {\cal P}\simeq N_0 \,\Phi\left(\frac{N}{N_0}, \frac{\ell}{\sqrt{4DT}}\right).
\end{equation}
Further, this first term alone correctly (and exactly) describes the extreme
cases of $N=N_0$ and $N=0$, where the Stirling formula does not hold.  In the
following we shall rederive this dominant term within the framework of
macroscopic fluctuation theory~\cite{Bertini,Jona}.  This derivation will
also yield the most likely density history of the system, a quantity that is
not readily accessible by microscopic theory.

\section{Macroscopic Fluctuation Theory}
\label{MFT}

\subsection{Basic Formalism}

The macroscopic fluctuation theory (MFT) was originally developed and
employed in the context of non-equilibrium steady states of diffusive lattice
gases~\cite{Bertini,Jona,Tailleur,Bunin} and subsequently extended to
non-stationary settings~\cite{D07,HEPG,DG2009b,KM_var,KMS,MS2013,MS2014}.
The MFT, and its extensions to reacting particle systems~\cite{EK,MS2011},
has proven to be versatile and efficient.  We briefly outline the
(Hamiltonian form of the) MFT equations and refer the reader to the above
references for details.  For independent diffusing particles, the particle
number density field $q(x,t)$ and the canonically conjugate ``momentum''
density field $p(x,t)$ obey Hamilton equations
\begin{subequations}
\begin{eqnarray}
\hspace{3 cm}
  \partial_t q &=& D\partial_{x}^2 q-2D\partial_x (q \partial_x p), \label{original1} \\
\hspace{3 cm}
  \partial_t p &=& - D\partial_{x}^2 p-D(\partial_x p)^2. \label{original2}
\end{eqnarray}
\end{subequations}
The Hamiltonian is $\int_0^{\infty} h \,dx$, where
$$
h=-D\partial_x p \,\partial_x q+D q(\partial_x p)^2.
$$
The boundary conditions at the absorber at $x=0$ are
$q(x\!=\!0,t)=p(x\!=\!0,t)=0$.  The boundary conditions in time are the
following. At $t=0$ we have
\begin{equation}\label{t0}
    q(x,t\!=\!0)=N_0 \delta (x-\ell).
\end{equation}
The condition
\begin{equation}\label{number}
\int_0^{\infty} q(x,T) \,dx = N
\end{equation}
imposes an integral constraint on the solution; this setting is similar to
that studied by Derrida and Gerschenfeld \cite{DG2009b}, see also
Refs.~\cite{KM_var,MS2013,MS2014}.  A derivation, similar to that presented
in Ref.~\cite{DG2009b}, leads to the following boundary condition for $p$ at
$t=T$:
\begin{equation}\label{pT}
    p(x,t\!=\!T)=\lambda \theta(x),
\end{equation}
where $\theta(x)$ is the Heaviside step function, and $\lambda$ is an a
priori unknown Lagrange multiplier that is ultimately set by
Eq.~\eqref{number}.

The solution of the MFT equations for $q(x,t)$ yields the most likely density
history of the system that we seek.  Once $q(x,t)$ and $p(x,t)$ are found,
one can calculate the action $S$, which yields ${\cal P}(N)$ up to a
pre-exponential factor:
\begin{equation}\label{actionmain}
-\ln {\cal P} \simeq S = D \int_0^T dt\, \int_0^{\infty} dx\, \left(p\partial_t q-h\right)  =D \int_0^T dt \int_{0}^{\infty} dx\,
q\, (\partial_x p)^2.
\end{equation}

\subsection{Hopf-Cole transformation and solution of the MFT problem}
For independent diffusing particles, the MFT problem is exactly soluble via
the Hopf-Cole transformation: a canonical transformation from $(q,p)$ to $Q=q
e^{-p}$ and $P=e^p$ \cite{EK,DG2009b,KMS}.  The generating function of this
canonical transformation is
\begin{equation}\label{genfun}
\int_0^{\infty} F(q,Q)\, dx=\int_0^{\infty} \left[q \ln (q/Q)-q\right]\, dx.
\end{equation}
The transformed Hamiltonian is $\int H\,dx$, where
$$
H=-D \,\partial_x P \,\partial_x Q\,.
$$
In the new variables, the Hamilton equations for $Q$ and $P$
are decoupled:
\begin{subequations}
\begin{eqnarray}
\hspace{3.7 cm}
  \partial_tQ &=& D \partial_x^2 Q\,, \label{Qt}\\
\hspace{3.7 cm}
  \partial_tP &=& - D \partial_x^2 P\,. \label{Pt}
\end{eqnarray}
\end{subequations}
As shown in \ref{app:action}, the action (\ref{actionmain}) can be written as
\begin{equation}\label{S}
    S\!=\!\int_0^{\infty}\!\!\! dx\left\{q(x,T)\Big[\ln\frac{q(x,T)}{Q(x,T)}\!-\!1\Big] -q(x,0)\Big[\ln\frac{q(x,0)}{Q(x,0)} \!-\!1\Big]\!+\!Q(x,T)\!-\!Q(x,0)\right\},
\end{equation}
which is fully determined by the initial and final states of the system.
However, to determine $Q(x,t)$ and $q(x,T)$, we need to find the entire
phase trajectory of the system.  Since Eqs.~(\ref{Qt}) and (\ref{Pt}) are
decoupled, we can solve the anti-diffusion equation (\ref{Pt}) backward in
time, with the initial condition $P(x,T)=1+(e^{\lambda}-1) \theta(x)$ and the
boundary conditions $P(0,t)=1$ and $P(\infty,t)=e^{\lambda}$.  The solution
is
\begin{equation}\label{stepP}
    P(x,t)=1+(e^{\lambda}-1)\, \text{erf}\left[\frac{x}{\sqrt{4D(T-t)}}\right].
\end{equation}
At $t=0$
we obtain
$$
Q(x,0)=\frac{q(x,0)}{P(x,0)}=\frac{N_0\,\delta(x-\ell)}{1+(e^{\lambda}-1)\,\text{erf}
\left(\frac{x}{\sqrt{4D T}}\right)} = \frac{N_0\,\delta(x-\ell)}{1+(e^{\lambda}-1)\, \text{erf}\left(\frac{\ell}{\sqrt{4 D T}}\right)}~.
$$
This expression serves as the initial condition for solving the diffusion
equation~(\ref{Qt}) forward in time with the boundary conditions
$Q(0,t)=q(0,t)/P(0,t)=0$ and $Q(\infty,t)=0$.  The solution is
\begin{equation}\label{Q(t)}
    Q(x,t)=\frac{N_0}{\sqrt{4\pi D t}}\,\frac{e^{-\frac{(x-\ell)^2}{4Dt}}-e^{-\frac{(x+\ell)^2}{4Dt}}}{1+(e^{\lambda}-1)\, \text{erf}\left(\frac{\ell}{\sqrt{4 D T}}\right)}~.
\end{equation}
We can now find $q(x,T)=Q(x,T) P(x,T)=e^{\lambda} Q(x,T)$ and
evaluate the action in Eq.~(\ref{S}):
\begin{equation}\label{Slambda}
    S=\frac{N_0 \lambda e^{\lambda} \overline{\Theta}}{1+(e^{\lambda}-1) \overline{\Theta}}
    -N_0 \log \left[1+(e^{\lambda}-1) \overline{\Theta}\right].
\end{equation}
Now we use Eq.~(\ref{number}) to express $\lambda$ via $N$:
$$
\frac{N_0  e^{\lambda} \,\overline{\Theta}}{1+(e^{\lambda}-1)\, \overline{\Theta}}=N,
$$
which yields
\begin{equation}\label{lambdaN}
    \lambda=\ln \left[\frac{\Theta (1-\overline{\Theta})}{(1-\Theta) \overline{\Theta}} \right].
\end{equation}
Substituting Eq.~(\ref{lambdaN}) into (\ref{Slambda}), we obtain
\begin{equation}\label{SN}
    -\ln {\cal P}\simeq S=N_0\left[\Theta \ln \frac{\Theta}{\overline{\Theta}}+(1-\Theta) \ln \frac{1-\Theta}{1-\overline{\Theta}}\right].
\end{equation}
This expression coincides with the leading term of Eq.~(\ref{Pappr}), which
was obtained from the exact solution (\ref{binom}) in the regime $N_0\gg 1$,
$N\gg 1$ and $N_0-N\gg 1$.  By normalizing the approximate distribution
(\ref{SN}) in the Gaussian region, we can also obtain the subleading term in Eq.~(\ref{Pappr}) [but with
$\overline{\Theta}(1-\overline{\Theta})$ instead of $\Theta(1-\Theta)$ inside the square root].

Now let us focus on the most likely density history, as described by
$q(x,t)=Q(x,t) P(x,t)$. Using Eqs.~(\ref{stepP}), (\ref{Q(t)}),
(\ref{lambdaN}) and (\ref{mf1}), we obtain
\begin{equation}\label{q}
   q(x,t)=
   \frac{\overline{\Theta} (1-\Theta)+(\Theta-\overline{\Theta}) \,\text{erf} \left[\frac{x}{\sqrt{4D(T-t)}}\right]}{\overline{\Theta}(1-\overline{\Theta})}\,\,q_{\text{mf}}(x,t),
\end{equation}
where
\begin{equation}\label{mfN0}
q_{\text{mf}}(x,t) = N_0 \,c(x,t)=\frac{N_0}{\sqrt{4\pi D t}}\left[e^{-\frac{(x-\ell)^2}{4Dt}}-e^{-\frac{(x+\ell)^2}{4Dt}}\right]
\end{equation}
gives the mean-field density history, which is not conditional on any number
of survivors at $t=T$.

Equation~(\ref{q}), together with \eqref{binom}, are the main results of this
work.  Of particular interest is the density profile at $t=T$:
\begin{equation}\label{q(T)}
   q(x,T)= \frac{N}{\overline{\Theta} \sqrt{4\pi D T}} \left[ e^{-\frac{(x-\ell)^2}{4DT}}-e^{-\frac{(x+\ell)^2}{4DT}}\right].
\end{equation}
By virtue of Eq.~(\ref{mfN0}), we obtain
\begin{equation}\label{q(T)simple}
   \frac{q(x,T)}{q_{\text{mf}}(x,T)} = \frac{N}{\overline{N}(T)}\,,
\end{equation}
where $\overline{N}(T)$ is given by Eq.~(\ref{barN}).  The most likely density
profile at $t=T$, conditional on $N$ particles surviving, differs from the
unconditional profile, where $\overline{N}(T)$ particles have survived, only by
the position-independent factor $N/\overline{N}(T)$.  For $t<T$, however, the two
density profiles are spatially quite dissimilar, as is evident from
Eq.~(\ref{q}).

A particularly useful characteristic of the survival history is the most
likely number of surviving particles $n(t)$ at intermediate times
$0<t<T$, conditioned on there being exactly $N$ survivors at $t=T$.  To
evaluate this quantity, we transform the particle flux in
Eq.~\eqref{original1} to the new variables $Q$ and $P$:
\begin{equation}
 \label{flux}
    j(x,t)=-D \partial_x q+2 D q \partial_x p =
D Q(x,t) \partial_x P(x,t)-D P(x,t) \partial_x Q(x,t)\,.
\end{equation}
Using Eqs.~(\ref{stepP}) and (\ref{Q(t)}) for $P$ and $Q$, we evaluate
$j(x\!=\!0,t)$ and integrate it over time from $0$ to $t$. The result is
\begin{equation}\label{survivors}
\frac{n(t)}{N_0}=1-\left(1-\frac{N}{N_0}\right)\,\frac{\text{erfc}\left(\frac{\ell}{\sqrt{4 D t}}\right)}
{\text{erfc}\left(\frac{\ell}{\sqrt{4 DT}}\right)}~,
\end{equation}
where $\text{erfc}(z) =1-\text{erf}(z)$.

\begin{figure}
\includegraphics[width=0.48\textwidth,clip=]{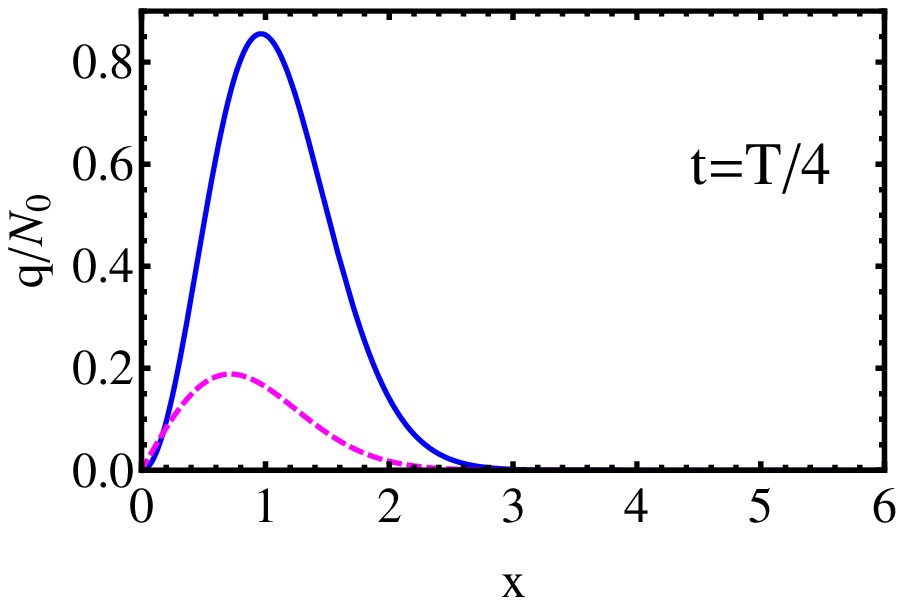}
\includegraphics[width=0.48\textwidth,clip=]{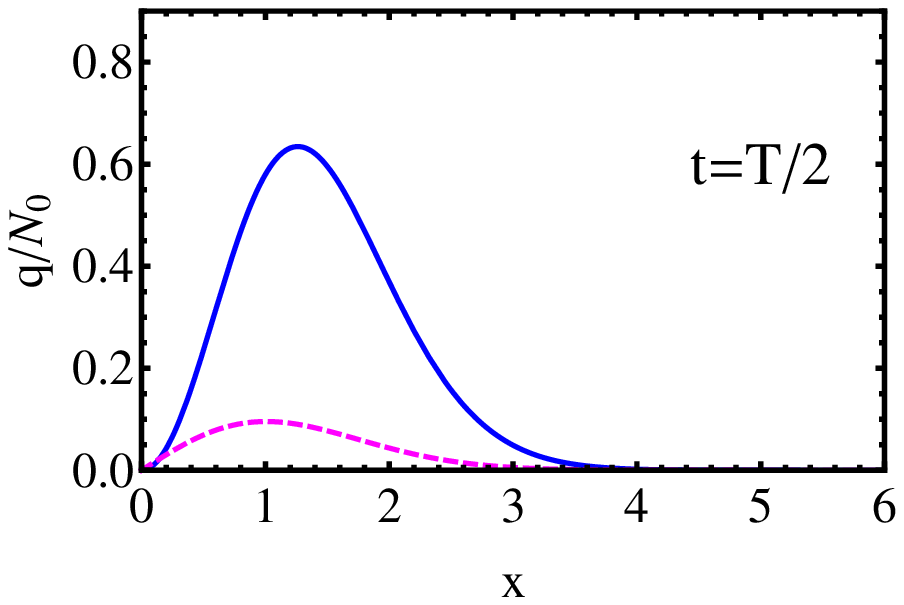}
\includegraphics[width=0.48\textwidth,clip=]{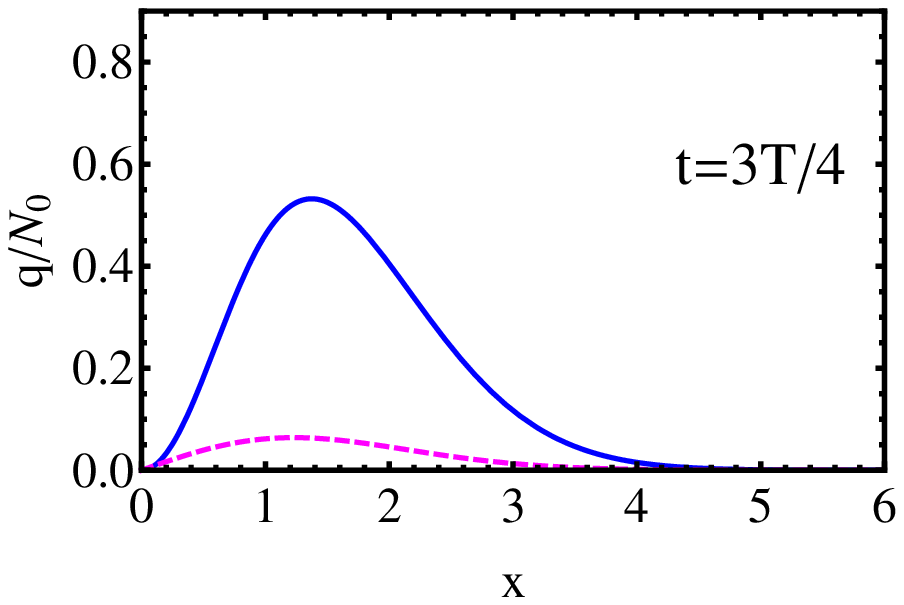}\hskip 0.24in
\includegraphics[width=0.48\textwidth,clip=]{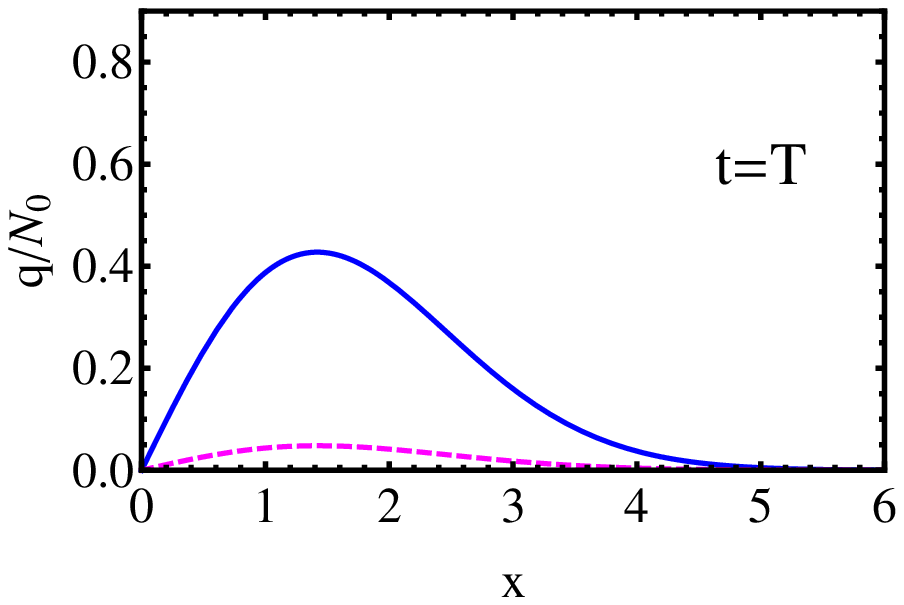}
\caption{The most likely density history, conditional on
  survival of all $N_0$ particles by time $t=T$, Eq.~(\ref{qall}) (the solid
  line). Also shown, by the dashed line, is the most likely unconditional
  density history, Eq.~(\ref{mfN0}). The parameters are $\ell=1/5$, $T=1$ and
  $D=1$.}
\label{history}
\end{figure}

Now let us consider the two extreme examples already discussed in
Sec.~\ref{P(N)}.  For the subset of histories where all the particles survive
up to time $T$, namely $N=N_0$, Eq.~\eqref{q} becomes
\begin{equation}\label{qall}
   q(x,t)= \frac{N_0}{\sqrt{4\pi D t}} \left[ e^{-\frac{(x-\ell)^2}{4Dt}}-e^{-\frac{(x+\ell)^2}{4Dt}}\right]\,
  \frac{\text{erf} \left[\frac{x}{\sqrt{4D(T-t)}}\right]}{\text{erf}\left(\frac{\ell}{\sqrt{4DT}}\right)}.
\end{equation}
In this case, the optimal fluctuation acts to make the particle flux
(\ref{flux}) zero at $x=0$ for all times $0<t<T$.
For $\ell/\sqrt{4DT}\ll 1$, $q(x,t)$ in
Eq.~(\ref{qall}) becomes independent of $\ell$:
\begin{equation*}
    q(x,t)\simeq \frac{N_0 \sqrt{DT}\, x \,e^{-\frac{x^2}{4 D t}}\,
   \text{erf}\,\left[\frac{x}{\sqrt{4D(T-t)}}\right]}{2 (Dt)^{3/2}}\,.
\end{equation*}
In this regime, the survival of all particles until time $T$ can be achieved
only as a result of a large fluctuation.  Figure~\ref{history} compares the
profiles of $q(x,t)$ and $q_{\text{mf}}(x,t)$ at different times for
$\ell/\sqrt{4DT}=1/10$.

\begin{figure}
\includegraphics[width=0.48\textwidth,clip=]{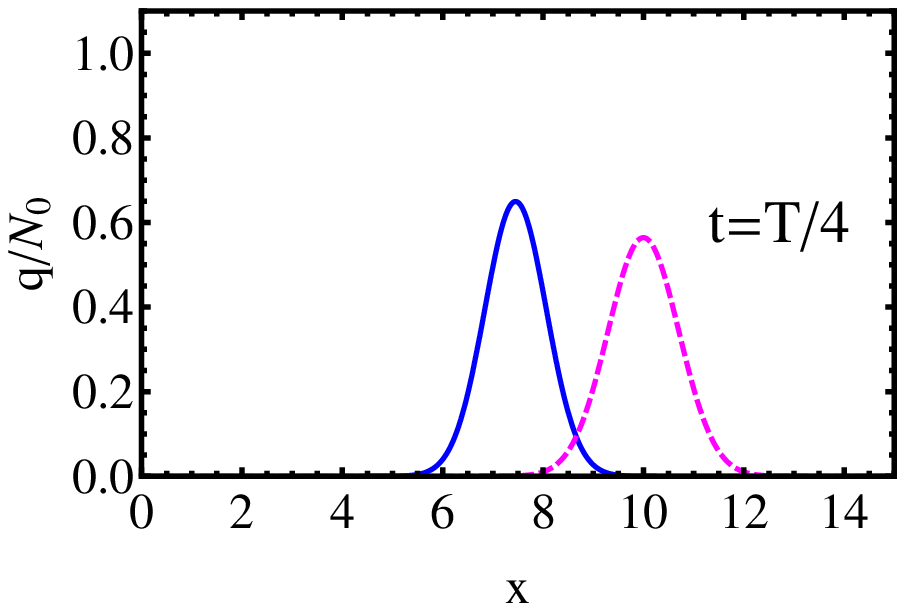}
\includegraphics[width=0.48\textwidth,clip=]{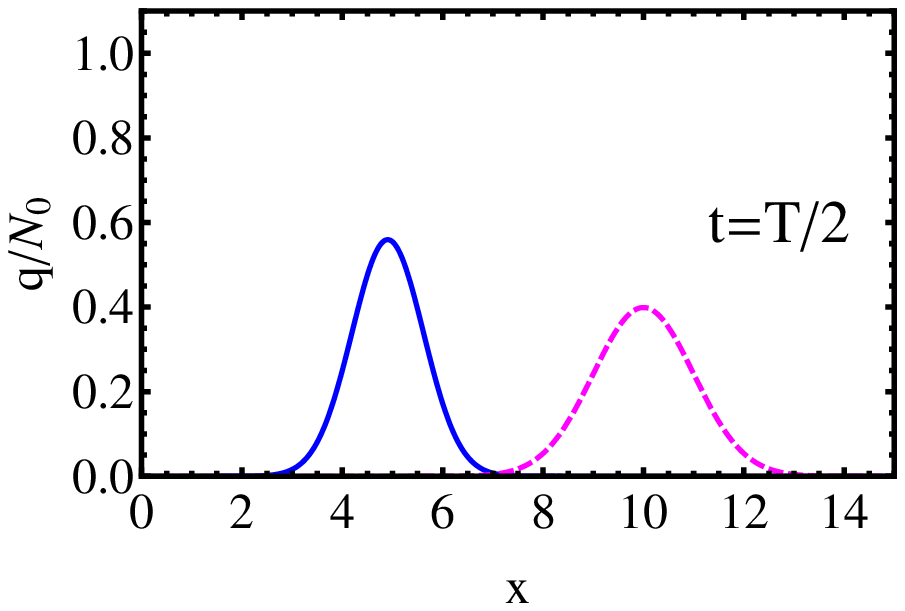}
\includegraphics[width=0.48\textwidth,clip=]{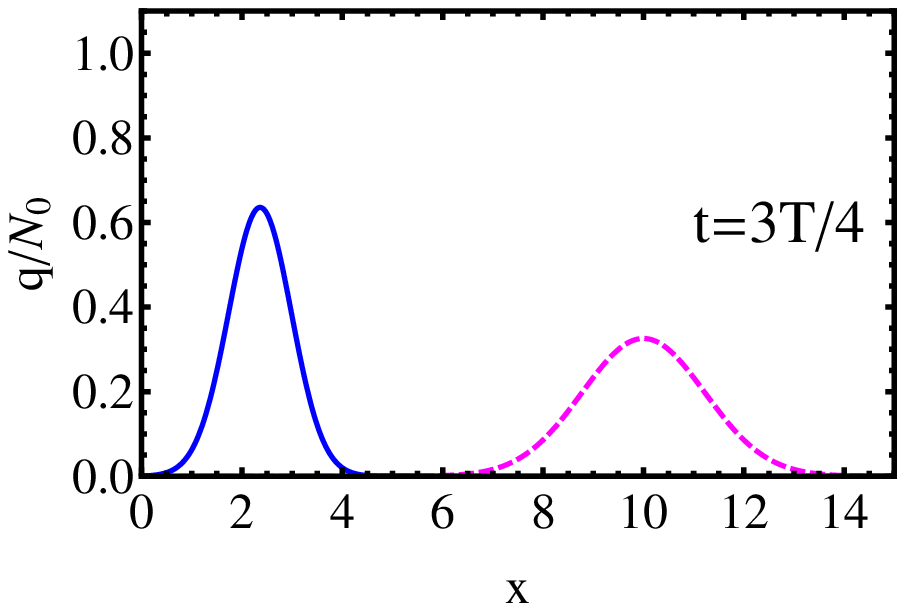}\hskip 0.24in
\includegraphics[width=0.48\textwidth,clip=]{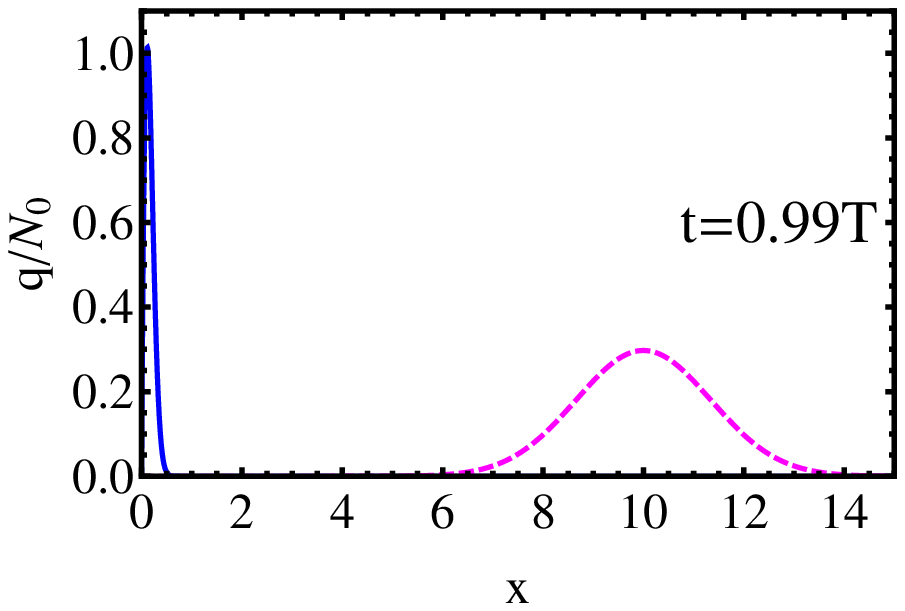}
\caption{The most likely density profile history, conditional on extinction
  of all $N_0$ particles by time $t=T$, Eq.~(\ref{qall}) (the solid
  line). Also shown, by the dashed line, the most likely unconditional
  density profile history, Eq.~(\ref{mfN0}). The parameters are $\ell=10$,
  $T=1$ and $D=1$.}
\label{history_ext}
\end{figure}

Even more interesting is the example of $N=0$, corresponding to no survivors
at $t=T$.  Now Eq.~(\ref{q}) becomes
\begin{equation}\label{qnone}
   q(x,t)= \frac{N_0}{\sqrt{4\pi D t}} \left[ e^{-\frac{(x-\ell)^2}{4Dt}}-e^{-\frac{(x+\ell)^2}{4Dt}}\right]\,
  \frac{\text{erfc} \left[\frac{x}{\sqrt{4D(T-t)}}\right]}{\text{erfc}\left(\frac{\ell}{\sqrt{4DT}}\right)}.
\end{equation}
In this case, it is the regime of $\ell/\sqrt{4DT}\gg 1$ that is controlled
by a large fluctuation.  Figure~\ref{history_ext} compares the profiles of
$q(x,t)$ and $q_{\text{mf}}(x,t)$ at different times for $\ell/\sqrt{4DT}=5$.
For this parameter value, the \emph{expected} number of survivors is only
slightly less than $N_0$.  Here the density profile, conditional on no
survivors at $t=T$, has the form of a relatively narrow pulse that moves
toward the absorber and is absorbed at $t\simeq T$.

We now make these qualitative observations quantitative.  For
$\ell/\sqrt{4DT}\gg 1$, we use the large-argument asymptotic of
$\text{erfc}\,(\ell/\sqrt{4DT})$ in Eq.~(\ref{qnone}), see
Eq. (\ref{largez}).  For not too small values of $x$, the large-argument
asymptotic can also be used for the erfc function in the numerator of
Eq.~(\ref{qnone}), and we can ignore the negligible second Gaussian in the
square brackets.  As a result,
\begin{align}
\label{q1}
 q(x,t)&\simeq \frac{N_0 \ell}{x}\,\sqrt{\frac{T-t}{4 \pi D \,t \,T}}
\exp\left[-\frac{(x-\ell)^2}{4 Dt}-\frac{x^2}{4 D
    (T-t)}+\frac{\ell^2}{4DT}\right] \nonumber \\
&= \frac{N_0 \ell}{x}\,\sqrt{\frac{T-t}{4 \pi D\,t\,T}}\,\exp\left\{-\frac{T}{4 D t (T-t)}\,\left[x-x_*(t)\right]^2\right\},
\end{align}
where
\begin{equation}\label{x*}
x_*(t)= \ell(1-t/T).
\end{equation}
The peak of this density pulse at $x\simeq x_*(t)$ moves ballistically toward
the absorber with speed $\ell/T$, while the maximum density is
$$
q_{\text{max}}(t) \simeq q_*[x_*(t),t]\simeq N_0\,\left[4 \pi D t (1-t/T)\right]^{-1/2}
$$
and the characteristic pulse width, $\Delta(t) = \sqrt{Dt(1-t/T)}$.  The
pulse height decreases and the pulse width increases with time until $t=T/2$
and vice versa for $t>T/2$.  The strong inequality $\ell/\sqrt{4DT}\gg 1$
guarantees that $\Delta \ll \ell$ at all times.  Thus we can approximately
replace $x$ by $x_*(t)$ in the prefactor of Eq.~(\ref{q1}).  This yields the
Gaussian density profile
\begin{equation}\label{gauss1}
q(x,t)\simeq \frac{N_0}{\sqrt{2\pi}\,\Delta(t)}\,e^{-\frac{(x-x_*)^2}{2\Delta^2(t)}},
\end{equation}
that is generally valid except very close to $t=T$.  To leading order, this solution describes
a ballistically moving constant-mass quasiparticle.  This ballistic motion
arises because of the constraint that a large number of particles must be
transported from $x=\ell$ to $x=0$ in a very short time.  In this noise-dominated regime
we can neglect the second
derivatives in the MFT equations (\ref{original1}) and (\ref{original2}) to
lowest order, leading to the reduced MFT equations~\cite{MS2014}
\begin{subequations}
\begin{eqnarray}
\hspace{3.5 cm}
  \partial_t q &=& -2D \partial_x (q \partial_x p), \label{invq} \\
\hspace{3.5 cm}
  \partial_t p &=& -D (\partial_x p)^2. \label{invp}
\end{eqnarray}
\end{subequations}
Now we need to solve Eq.~(\ref{invp}) backward in time with the initial
condition (\ref{pT}).  In view of Eq.~(\ref{lambdaN}), the condition $N=0$
corresponds to $\lambda=-\infty$.  That is, $p(x\!>\!0,T) \to -\infty$ as $t
\to T$.  The appropriate solution is (cf.\ Ref.~\cite{MS2014})
\begin{equation}\label{vRW}
    \partial_x p(x>0,t) = -\frac{x}{2D (T-t)}\,.
\end{equation}
Equation~(\ref{invq}) is a continuity equation for the density $q(x,t)$ with
velocity field $2D\, \partial_x p$, with $\partial_x p$ determined from
Eq.~(\ref{vRW}).  As one can easily check, its exact (generalized, or weak)
solution for the delta-function initial condition (\ref{t0}) is the
translating delta function:
\begin{equation}\label{traveldelta}
q(x,t)=N_0 \,\delta[x-x_*(t)]\,,
\end{equation}
with $x_*(t)$ given by Eq.~(\ref{x*}).  In this limit, the quasiparticle is
simply a material point.  Upon its release at $x=\ell$, this point moves
ballistically with speed $\ell/T$ until it hits the absorber.  Let us
calculate the quasiparticle contribution to the action, using
Eq.~(\ref{actionmain}):
\begin{align}
\label{sinviscid}
-\ln {\cal P}(N\!=\!0)= S &= D\! \int_0^T \!\! dt \int_{0}^\infty \!\!dx
\,q\,(\partial_x p)^2\nonumber \\
&=D \int_0^T \!\! dt \int_{0}^{\infty}\!\! dx \,N_0\,\delta[x-x_*(t)]\,\frac{x^2}{4D^2 (T\!-\!t)^2}
  = \frac{N_0 \ell^2}{4DT}.
\end{align}
This result coincides with the leading-order term in Eq.~(\ref{0large}). That
is, the dominant contribution to the action comes from a ballistically moving
quasiparticle.  A similar effect is observed when an unusually large mass or
energy is transported in a short time in \emph{interacting} diffusive lattice gases of the
so-called hyperbolic class \cite{MS2013}.

\begin{figure}
\centerline{\subfigure[]{\includegraphics[width=0.45\textwidth,clip=]{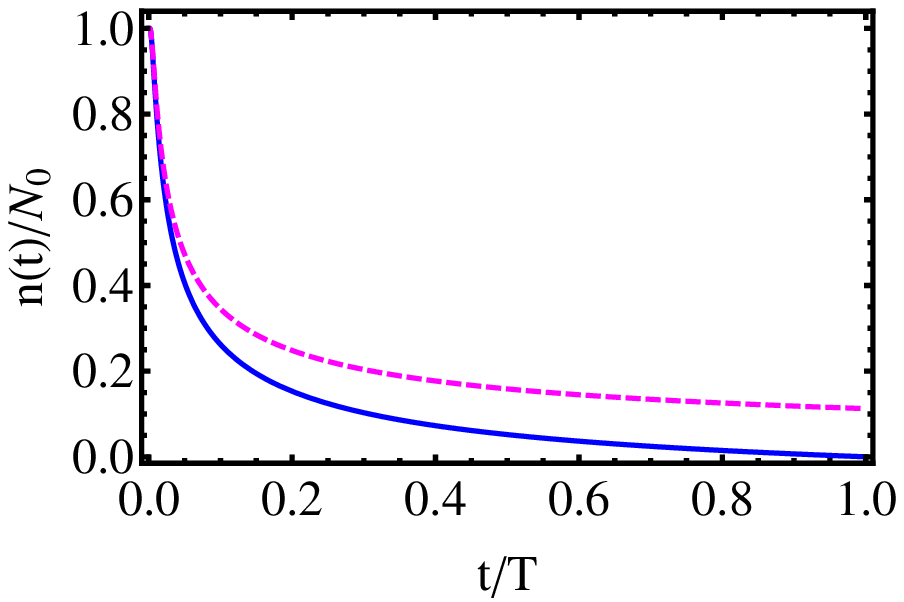}}\qquad\qquad
\subfigure[]{\includegraphics[width=0.45\textwidth,clip=]{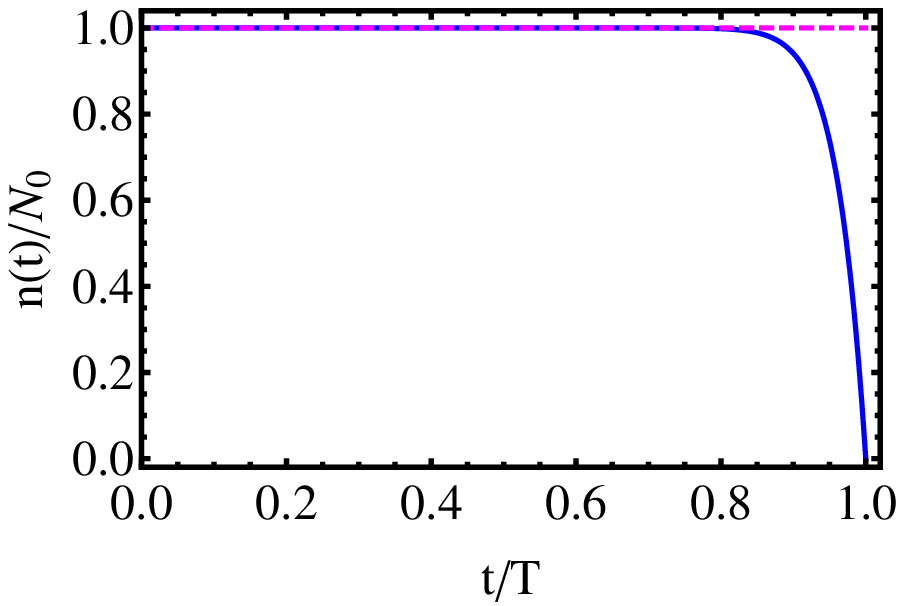}}}
\caption{Solid line: the fraction of survivors $n(t)/N_0$ versus $t/T$,
  conditional on no survivors at $t=T$ for (a) $\ell/\sqrt{4DT}=1/10$ and (b)
  $\ell/\sqrt{4DT}=5$.  Dashed line: the most likely unconditional fraction
  of survivors $\overline{\Theta}(t)$ from Eq.~(\ref{expectedSresult}).}
\label{survivorsfig}
\end{figure}

Finally, let us examine how the most probable number of particles $n(t)$ in
Eq.~(\ref{survivors}) depends on time under the constraint that no
particles survive at time $t=T$.  Figure \ref{survivorsfig} shows the
fraction of survivors $n(t)/N_0$ versus $t$ for: (a)
$\ell/\sqrt{4DT}=1/10$ and (b) $5$.  In the former case, most of the
particles would get absorbed by $t=T$ ``naturally'', without need for large
fluctuations.  In the latter case almost all particles would typically
survive up to $t=T$, so a large fluctuation is needed to ensure that all of
them are absorbed.  The latter case corresponds to the evolution illustrated
in Fig.~\ref{history_ext}, where a quasiparticle moves ballistically toward
the absorber so that essentially all particles are absorbed when $t=T$.

\section{Concluding Remarks}
\label{discussion}

We investigated the role of large fluctuations in diffusion-controlled
absorption in one dimension.  We determined the probability distribution of
the number of particles $N$ that have not been absorbed by time $T$.  Apart
from $N$ and $N_0$, this distribution crucially depends on the parameter
$\ell/\sqrt{4DT}$.

We employed the macroscopic fluctuation theory (MFT) to find the ``optimal
path'' of the system, namely, the most probable density history conditioned
on a given number of surviving particles at an arbitrary observation time
$t=T$.  The optimal path gives fascinating insights into the nature of large
fluctuations of the absorption process.  A striking result arises in the
situation where one demands that the particles, released far from the
absorber, are absorbed in time $T$ that is short.  Here, to leading order,
the spatial probability density moves like a material point particle with
constant speed towards the origin, and is absorbed at time $T$.

The model of non-interacting diffusing particles is amenable to a complete
analytical solution within the framework of the MFT~\cite{EK,DG2009b,KMS}.
For diffusive lattice gases of interacting particles, the MFT problem becomes much
harder to solve. Nevertheless, some of the insights we gained here should be
useful in studying particle or energy absorption and other dynamical
processes in interacting lattice gases.

\medskip

We thank Paul Krapivsky for helpful discussions.  Financial support of this
research was provided in part by BSF grant No.\ 2012145 (BM and SR) and NSF
Grant No.\ DMR-1205797 (SR).

\appendix
\section{Action}
\label{app:action}

The action can be written as
\begin{eqnarray}
  S &=& \int_0^T dt \int_0^{\infty} dx \left(p\partial_t q-w\right)  \nonumber \\
  &=&\int dx\,F(q,Q) \biggl|_0^T +\int_0^T dt \int_0^{\infty} dx \left(P\partial_t Q+D \partial_xP \partial_xQ\right) \nonumber \\
  &=& \int dx\, F(q,Q) \biggl|_0^T+
\int_0^T dt \int_0^{\infty} dx  \left(D P\partial_x^2 Q+D \partial_xP\partial_xQ\right) \nonumber \\
  &=&\int dx\, F(q,Q) \biggl|_0^T-\int_0^T \,dt P(0,t) D \partial_x Q(0,t). \label{algebra1}
\end{eqnarray}
As $p(0,t)=0$, we have $P(0,t)=1$. Using Eq.~(\ref{Qt}), we obtain
$$
-D \partial_x Q(0,t)=\int_0^{\infty} \partial_t Q\,dx.
$$
Therefore, the last term in Eq.~(\ref{algebra1}) can be written as
$$
-\int_0^T \,dt \,P(0,t) D \partial_x Q(x,t)=\int_0^Tdt\,\partial_t \int_0^{\infty} Q\,dx =
\int_0^{\infty}dx\, Q(x,t) \biggl|_0^T,
$$
and we obtain Eq.~(\ref{S}) for the action.

\end{document}